\documentclass[lettersize,journal]{IEEEtran}
\IEEEoverridecommandlockouts
\usepackage{cite}
\usepackage{amsmath,amssymb,amsfonts}
\usepackage{graphicx}
\usepackage{textcomp}
\usepackage{xcolor}
\usepackage{mathptmx}
\usepackage{lipsum}
\usepackage{hyperref}
\usepackage{url}
\usepackage{algpseudocode}
\usepackage{svg}
\usepackage{booktabs}
\usepackage{algorithm}
\usepackage{microtype}
\usepackage{fixltx2e}
\newif\ifshowcomments
\showcommentsfalse

\definecolor{harry}{RGB}{0, 128, 192}

\def\BibTeX{{\rm B\kern-.05em{\sc i\kern-.025em b}\kern-.08em
    T\kern-.1667em\lower.7ex\hbox{E}\kern-.125emX}}
\begin{document}

\title{Targeted Power System Frequency Attack via the Selection of Maliciously Controlled Inverters
}

\author{Xiangyu Zou, Betelihem Ashebo, and Daniel K. Molzahn

\thanks{X. Zou, B. Ashebo, D.K. Molzahn are with the School of Electrical and Computer Engineering, Georgia Institute of Technology, Atlanta, GA, USA. Email: \{\href{mailto:xy.zou@gatech.edu}{xy.zou}, \href{mailto:bashebo3@gatech.edu}{bashebo3}, \href{mailto:Molzahn@gatech.edu}{molzahn}\}@gatech.edu.}
\thanks{Support from the U.S. Department of Energy, Office of Science, Office of Cybersecurity, Energy Security, and Emergency Response (CESER), under DE-CR0000056 and the National Science Foundation under grant number \#2112533: NSF Artificial Intelligence Research Institute for Advances in Optimization (AI4OPT).}}

\maketitle

\begin{abstract}
This paper studies how an adversary can execute a power system frequency attack by choosing the most effective subset of inverter-based resources (IBRs) as malicious control nodes. During the attack, the adversary controls the attacking devices to destabilize a group of designated target generators. The attack is designed by introducing an unstable oscillatory mode whose eigenvector has large components at the target generators and small components at the compromised IBRs. We formalize the optimal attacker selection problem and present an equivalent mixed-integer quadratically constrained program (MIQCP). To address this combinatorial nonconvex problem, we develop two heuristic algorithms by introducing a ranking metric. The attack process is demonstrated on a modified WSCC 179-bus system, with results verified through nonlinear dynamic simulations. We show that the proposed heuristics find optimal attacking sets in a majority of evaluated scenarios while significantly reducing the required computational time. Scenarios from the \mbox{ACTIVSg500} system are used to further support our results. Finally, we discuss how selecting different eigenvalues, targets, and numbers of compromised devices impact the attack's severity. 
\end{abstract}

\begin{IEEEkeywords}
Cyber-physical security, inverter-based resources, stability, malicious control, mixed-integer programming.
\end{IEEEkeywords}

\section{Introduction }\label{Sec1:Intro}

\IEEEPARstart{T}{he} power grid is undergoing a major transformation with
the widespread integration of inverter-based resources (IBRs), including
solar photovoltaic (PV), wind generators, and energy storage systems. Renewable sources accounted for 92.5\% of new global power generation capacity in 2024, up from 85.8\% in 2023~\cite{irena_digitalisation_ai_2025}, with solar PV and wind accounting for 96.6\% of renewable additions~\cite{irena_capacity_highlights_2025}. This expansion is expected to continue, and is projected to account for approximately 95\% of global renewable capacity additions through 2030~\cite{iea_renewables_2024}. IBRs interface with the grid through power electronics governed by digital control systems. This shift fundamentally alters power system dynamics by replacing the electromechanical response of synchronous generators with the control-determined behavior of converter-interfaced devices.

IBRs introduce distinct cybersecurity vulnerabilities, as their reliance on digital controllers, communication interfaces, and configurable software settings significantly expands the attack surface~\cite{nrel2025ibr_cyberphysical, osti_1832209}. Adversaries with device-level access could manipulate measurements, control logic, or setpoints, potentially causing frequency instability, voltage violations, or thermal overloading~\cite{tuyen22,ahn24}. 

A coordinated cyberattack involving multiple compromised IBRs would provide an adversary with a wider range of control actions than a single-device breach. However, compromising additional devices requires additional time, access, and resources. An attacker must therefore determine which IBRs will have the greatest grid impact when compromised and allocate resources accordingly. This work characterizes such worst-case attacks by identifying the devices most capable of destabilizing the system. Identifying these high-impact resources is important for assessing the potential consequences of coordinated attacks and prioritizing the deployment of cybersecurity protections.

Previous research has shown that coordinated malicious control can excite unstable
modes that target selected generators~\cite{demarco98,roberts21}. However, these studies assume that the set of compromised devices is fixed and known \emph{a priori}. Allowing the adversary to select the compromised IBRs adds an additional degree of freedom. The effectiveness of an attack may therefore depend not only on \emph{how} the compromised resources are controlled, but also \emph{which} devices are under malicious control.

Accordingly, this paper introduces the problem of selecting a subset of IBRs that an adversary would prioritize to maximize the destabilizing effect on a designated set of target synchronous generators. We analyze this device-selection problem using a small-signal stability framework, which characterizes local system dynamics by linearizing the governing equations around an operating point \cite{anderson03,saadat10}. The attack is designed using an eigenvalue and eigenvector placement approach and cast into a combinatorial optimization problem.

Next, we review the literature on DER cybersecurity, small-signal analysis, malicious control, and control node selection. 

\subsection{Related work}
The dependence of IBRs on digital controllers, communication links, and configurable software settings makes their control systems vulnerable to remote compromise. Surveys of the IBR attack surface describe how communication and firmware vulnerabilities may allow unauthorized changes to control parameters, setpoints, and reference signals~\cite{tuyen22,ahn24}. Specific inverter control functions have also been examined. Malicious alteration of an inverter's volt-var control curve can degrade system stability and performance~\cite{saber23}. Similarly, manipulation of the phase-locked loops of grid-following inverters can force them into off-nominal frequency operation and disturb the surrounding system~\cite{bamigbade2023cyberattack}. Reliability authorities have also identified IBR control and protection settings as contributors to large-scale disturbances, underscoring the consequences of incorrect or unauthorized changes to these settings~\cite{nerc_ibr_report_2023}. These studies show that unauthorized modification of inverter control functions is technically plausible. 

The dynamic behavior exploited by these attacks can be studied through small-signal stability analysis. In this framework, the nonlinear equations governing system dynamics are linearized around a steady-state operating point, and the eigenvalues of the resulting system matrix determine the stability and damping of its oscillatory modes~\cite{anderson03,saadat10}. With increasing IBR penetration, small-signal models have been extended to represent fast converter control loops~\cite{he21}. IBRs commonly operate under active and reactive power droop control, which relates frequency and voltage deviations to changes in power output~\cite{bahrani24}. The associated droop gains and time constants influence system eigenvalues and may alter modal damping and coupling among IBRs~\cite{yang23}.

Stability analysis is generally used to identify or mitigate unstable behavior. In contrast, malicious control seeks to modify the same modal structure to create instability. This idea was introduced in \cite{demarco96,demarco98}, where a group of generators applies feedback control to excite an unstable mode in which designated target generators have the largest participation. The targets are driven toward their protection limits while the attacking units remain connected. Later studies considered similar attacks using demand-side loads under proportional control~\cite{amini15}, aggregated loads providing emulated inertia~\cite{brown18}, and active loads in islanded microgrids~\cite{roberts21}. The system and generator information needed to construct such attacks may be obtained through insider access or information leakage. Reference~\cite{brown18} also showed that the attack can remain effective under uncertainty in generator parameters and network topology. The work in~\cite{amini2016dynamic} optimizes load perturbations to move selected eigenvalues into the right-half plane, showing that deliberate destabilization can be treated as a control-design problem.

The studies discussed above assume that the compromised devices are fixed in advance, that is, the attacker is assumed to already control a particular generator, load, or inverter. In contrast, the problem considered here allows the adversary to choose which IBRs to compromise. This additional decision creates a combinatorial device-selection problem that has not been addressed in the existing malicious-control literature.

The problem is structurally related to input and actuator selection in systems control. Input/output selection determines where the inputs and outputs of a system should be placed to achieve goals of controllability and disturbance rejection~\cite{vandewal01}. Actuator-placement methods similarly select a limited set of control inputs, often through mixed-integer optimization, to achieve controllability while minimizing control effort~\cite{chanekar17}. Such selection problems are computationally difficult in general~\cite{olshevsky2014minimal}, and has motivated the use of controllability metrics and submodular greedy methods for tractable actuator selection~\cite{summers2015submodularity}.

These formulations, however, are designed to improve the ability to steer the system over a broad range of states. The adversary considered in this paper has a more specific objective and does not require full controllability. The goal is to introduce a single unstable mode with large modal participation at the designated target generators and limited participation at the compromised IBRs.
Reference~\cite{katewa2021optimal} provides a related formulation in which load perturbations are selected to destabilize the system through eigenvalue placement. The present problem differs by directly optimizing the relative modal participation of the target and attacking resources so that the induced instability primarily affects the intended targets while the compromised IBRs remain connected and thus capable of initiating further attacks or benefiting economically. To the best of our knowledge, this objective has not previously been posed as a device-selection problem. Section~\ref{Sec3:Formulation} formulates it through the direct design of the closed-loop eigenstructure.

\subsection{Proposed Framework and Contributions}

This paper extends previous research on malicious control in power systems \cite{demarco96,demarco98,roberts21} by formalizing the optimal cyber-physical attacker selection problem. We identify the set of devices that, when acting maliciously, has the greatest destabilizing effect on a predefined set of target generators.
Several algorithms are proposed to solve the combinatorial problem, including exhaustive search, a mixed-integer optimization framework, and two ranking-based heuristics. We show that the problem greatly simplifies for certain scenarios, such as when targeting a single generator. We also demonstrate how the frequency the attack operates on, represented by a prescribed eigenvalue, impacts the attack's effectiveness.

We illustrate our methods via case studies using the Western Systems Coordination Council (WSCC) 179-bus and the \mbox{ACTIVSg500} test cases~\cite{zou2026}, showing through nonlinear dynamical simulations that the compromised devices inflict unstable oscillations in the target generators. 

In summary, the contributions of this work are as follows:
\begin{enumerate}
    \item A formal optimization-based representation of the optimal attacker selection problem for inverter-based resources, where the objective is to identify a subset of devices that produces the largest destabilizing effect on specified target generators.
    
    \item A mixed-integer quadratically-constrained program (MIQCP) reformulation that can be solved by state-of-the-art solvers such as Gurobi~\cite{gurobi23}. 
    
    \item Two ranking-based heuristics to circumvent the combinatorial complexity of the optimization problem, with numerical results to support their effectiveness.

    \item Intuition into how the selection of the attack's frequency influences the problem as well as how certain special cases lead to simplifications in the problem's structure.
\end{enumerate}

This paper is organized as follows. Section~\ref{Sec2:Prob} introduces the dynamical network model and its linearization. Section~\ref{Sec3:Formulation} formalizes attacker selection as a nonlinear optimization problem and presents an equivalent mixed-integer quadratically constrained program (MIQCP) form. Section~\ref{Sec4:heuristic} develops a heuristic search algorithm as a computationally lighter alternative. Section~\ref{Sec5:Case-Study} demonstrates the performance of the algorithms using an example of a successful attack. Section~\ref{Sec6:Conclusion} concludes the discussion with avenues for future research.

\section{Dynamic Model}\label{Sec2:Prob}

Section~\ref{Sub2a:dynamics} presents the nonlinear model of coupled generator and IBR dynamics. Section~\ref{Sub2b:linearize} linearizes this model around a power flow solution to enable eigenvalue analysis tools used later in the paper.

\subsection{Dynamic System Model} \label{Sub2a:dynamics}

    In a power system with $b$ buses, $g$ synchronous generators and $h$ inverter-based resources, the system's response to perturbations is governed by the devices' dynamic behaviors, which are coupled by the network's power flow equations.

    We adopt the one-axis (flux-decay) synchronous machine model from~\cite{sauer98}:
\begin{subequations} \label{eq-s2fluxdecay}
\begin{align} 
    \dot{\delta_i} &= \omega_i - \omega_s, \label{eq-s2fluxdecayA}\\
    \dot{\omega}_i &= \frac{\omega_s}{2H_i}(T_{M,i} + \Delta T_{M,i} - E_{q,i}' I_{q,i} \nonumber \\
    & \quad - (x_{q,i}-x_{d,i}')I_{d,i}I_{q,i} - D_i (\omega_i-\omega_s)), \label{eq-s2fluxdecayB}\\
    \dot{E}'_{q,i} &= \frac{1}{T_{do,i}'}(-E_{q,i}' - (x_{d,i}-x_{d,i}')I_{d,i} + E_{Fd,i}), \label{eq-s2fluxdecayC}
\end{align}    
\end{subequations}
for all $i \in \{1,\dots,g\}$. 
The machines have reactance parameters~$x_d$, $x_d'$, and $x_q$, generator inertia and damping parameters~$H$ and~$D$, \textit{d-q} frame current variables~$I_d$ and~$I_q$, internal voltage~$E_{q,i}'$, and field voltage~$E_{Fd}$. In \eqref{eq-s2fluxdecayA}, the rotor angle~$\delta$ varies when the frequency $\omega$ deviates from its nominal value~$\omega_{\rm s}$. Equation~\eqref{eq-s2fluxdecayB} models the variation in frequency when the mechanical torque $T_M$ and electrical torque represented by the~$I_q$ terms are not balanced. Finally, \eqref{eq-s2fluxdecayC} defines the dynamics of the internal voltage~$E_q'$. 

Each generator is equipped with an exciter that regulates the terminal voltage $V_i$ by adjusting the field voltage $E_{Fd}$ in~\eqref{eq-s2fluxdecayC}, with equations
    \begin{subequations} \label{eq-s2exciter}
    \begin{align}
        \dot{E}_{Fd,i} &= -\frac{1}{T_{E,i}}(K_{E,i} + S_{E,i}(E_{Fd,i})) E_{Fd,i} + \frac{V_{R,i}}{T_{E,i}}, \label{eq-s2exciterA} \\
        \dot{R}_{F,i} &= -\frac{1}{T_{F,i}} R_{F,i} + \frac{K_{F,i}}{T_{F,i}^2} E_{Fd,i}, \label{eq-s2exciterB} \\
        \dot{V}_{R,i} &= -\frac{V_{R,i}}{T_{A,i}} + \frac{K_{A,i}}{T_{A,i}} R_{F,i} - \frac{K_{A,i}K_{F,i}}{T_{A,i}T_{F,i}} E_{Fd,i} + \frac{K_{A,i}}{T_{A,i}}(V_{ref,i} - V_i), \label{eq-s2exciterC}
    \end{align}    
    \end{subequations}
for all $i \in \{1,\dots,g\}$. 
The stabilizer signal $R_F$ and voltage regulator output $V_R$ are defined by \eqref{eq-s2exciterB} and \eqref{eq-s2exciterC}, respectively. These adjust the field voltage in the exciter equation \eqref{eq-s2exciterA}.
The constants $K_A, K_E, K_F$ and $T_A, T_E, T_F$ correspond to control gains and time delays, respectively. The function $S_E(\cdot)$ models exciter saturation, and $V_{ref}$ is the generator's voltage reference.

Finally, the governor and steam-turbine models that control the mechanical torque $\Delta T_M$ in \eqref{eq-s2fluxdecayB} are described by
    \begin{subequations} \label{eq-s2governor}
    \begin{align}
        \Delta \dot{T}_{M,i} &= \frac{1}{T_{CH,i}}(-\Delta T_{M,i} + \Delta P_{SV,i}), \label{eq-s2governorA}\\
        \Delta \dot{P}_{SV,i} &= \frac{1}{T_{SV,i}}\left(-\Delta P_{SV,i} - \frac{1}{R_{D,i}\omega_s}(\omega_i - \omega_s)\right),  \label{eq-s2governorB}
    \end{align}
    \end{subequations}
for all $i \in \{1,\dots,g\}$. 
Equation~\eqref{eq-s2governorB} controls the steam valve signal $\Delta P_{SV}$, which is an input to the first-order turbine model in~\eqref{eq-s2governorA}. The quantities $T_{CH}$ and $T_{SV}$ are time constants, and $R_D$ is the inverse droop constant.

The system's IBRs use $P$-$f$ droop control, where each device adjusts its active power output when the frequency of the grid deviates from its nominal value:
    \begin{subequations} \label{eq-s2ibrdroop}
    \begin{align}
        \dot{\delta}_{H,i} &= \Delta \omega_{H,i},  \label{eq-s2ibrdroopA}\\
        \Delta \dot{\omega}_{H,i} &= \frac{1}{T_{pf}}(-\Delta \omega_{H,i} - m_{p,i} \omega_s \gamma_{H,i}(P_{H,i} - P_{ref,i})),  \label{eq-s2ibrdroopB}
    \end{align}    
    \end{subequations}
for $i \in \{1,\dots,h\}$. The droop law \eqref{eq-s2ibrdroopB} controls the IBR's internal angle $\delta_H$ which indirectly controls its power injection. The IBR has a time constant $T_{pf}$ and droop constant $m_p$. The quantity $\gamma_H$ is a unit conversion factor.

In total, there are $8g + 2h$ state variables. Furthermore, the generator currents $I_{d,i}$ and $I_{q,i}$ as well as voltage magnitudes~$V_i$ and phase angles~$\theta_i$ of all buses form $2g +2b$ algebraic variables that couple the ODEs. The stator algebraic equations
\begin{align} \label{eq-s2stator}
    V_i \sin(\delta_i - \theta_i) - x_{q,i} I_{q,i} &= 0, \nonumber \\
    E_{q,i}' - V_i \cos(\delta_i - \theta_i) - x_{d,i}' I_{d,i} &= 0,
\end{align}
and nodal power balance equations
\begin{align} \label{eq-s2network}
    P_{net,i} - \sum_{k=1}^b V_iV_kY_{ik} \cos(\theta_i-\theta_k-\alpha_{ik}) &= 0, \nonumber\\
    Q_{net,i} - \sum_{k=1}^b V_iV_kY_{ik} \sin(\theta_i-\theta_k-\alpha_{ik}) &= 0,
\end{align}
for all $i \in \{1,\dots,b\}$, together form the algebraic equations. The ($i,k$)-element in the nodal admittance matrix has magnitude $Y_{ik}$ and angle $\alpha_{ik}$. The net active and reactive power injections $P_{net,i}$ and $Q_{net,i}$ depend on which devices reside at the bus. The power injections from synchronous generators are defined by
\begin{align*}
    P_{G,i} &= I_{d,i} V_{G(i)} \sin(\delta_i - \theta_{G(i)}) + I_{q,i} V_{G(i)} \cos(\delta_i - \theta_{G(i)}), \\
    Q_{G,i} &= I_{d,i} V_{G(i)} \cos(\delta_i - \theta_{G(i)}) - I_{q,i} V_{G(i)} \sin(\delta_i - \theta_{G(i)}),
\end{align*}
and the power injections from IBRs are defined by
\begin{align*}
    P_{H,i} &= \frac{E_iV_{H(i)}}{x_{l,i}} \sin(\delta_{H,i}-  \theta_{H(i)}), \\
    Q_{H,i} &= \frac{V_{H(i)}^2 - E_iV_{H(i)}\cos(\delta_H - \theta_{H(i)})}{x_{l,i}}.
\end{align*}
where the indexing subscripts $G(i)$ and $H(i)$ denote the bus number of the $i$-th synchronous generator and IBR, respectively, and $E_i$ is the internal voltage of the $i$-th IBR. Active and reactive demand are treated as constant power loads.

Together, the ODEs~\eqref{eq-s2fluxdecay}--\eqref{eq-s2ibrdroop} and algebraic equations~\eqref{eq-s2stator},~\eqref{eq-s2network} form the system of differential-algebraic equations (DAE) that describes the nonlinear power system dynamics.

\subsection{Linearized System Model} \label{Sub2b:linearize}

Linearizing the DAE model~\eqref{eq-s2fluxdecay}--\eqref{eq-s2network} around a power flow solution yields a local approximation of the system dynamics. First, the initial values of all dynamic and algebraic variables of the DAE are found. Then, new variables are defined to represent the deviation from the initial steady-state operating point. 
Let $x \in \mathbb{R}^{8g+2h}$ and $y \in \mathbb{R}^{2g+2b}$ represent the changes in state and algebraic variables from their equilibrium values, respectively. By computing a first-order approximation of the dynamical equations, a linear system of the form
\begin{equation*}
    \begin{bmatrix} \dot{x} \\ 0 \end{bmatrix} = 
    \begin{bmatrix} \mathbf{A}_{11} & \mathbf{A}_{12} \\ \mathbf{A}_{21} & \mathbf{A}_{22} \end{bmatrix} 
    \begin{bmatrix} x \\ y \end{bmatrix}
\end{equation*}
is obtained, where the block matrices $\mathbf{A}_{11}, \mathbf{A}_{12}$ and $\mathbf{A}_{21}, \mathbf{A}_{22}$ contain coefficients of the linearized differential equations ~\eqref{eq-s2fluxdecay}--\eqref{eq-s2ibrdroop} 
and algebraic equations \eqref{eq-s2stator} and \eqref{eq-s2network}. The algebraic variables can be eliminated by Kron reduction (Schur complement)~\cite{anderson03} to produce the linear system
\begin{equation*}
    \dot{x} = \mathbf{A} x
\end{equation*}
where
\begin{equation*}
    \mathbf{A} = \mathbf{A}_{11} - \mathbf{A}_{12} \mathbf{A}_{22}^{-1} \mathbf{A}_{21}
\end{equation*}
is the size $n = 8g+2h$ square matrix that describes the system's dynamics around the power flow solution.

The stability of the linearized system to small disturbances is determined by the eigenvalues of $\mathbf{A}$. We focus on initially stable systems, i.e., all eigenvalues of $\mathbf{A}$ have negative real-parts. The goal of the adversary is to destabilize the system by selecting which devices to compromise and devising a malicious control strategy. The next section formalizes this goal more precisely and develops an optimization program for this process.
\section{Optimization Formulation} \label{Sec3:Formulation}

Section~\ref{Sub3a:threat} first describes an overview of the threat model, including the type and capability of the attack and other assumptions. Section~\ref{Sub3b:opt} then develops an optimization program to represent the attack selection and malicious control formulation process. 
Despite the nonconvex and combinatorial nature of the problem, Section~\ref{Sub3c:MIQCP} next shows an equivalent MIQCP form that is conducive to solvers such as Gurobi~\cite{gurobi23}.

\subsection{Malicious Control Threat Model} \label{Sub3a:threat}

We focus on attacks where the malicious actor gains control over several IBRs, referred to as attackers, with the goal of destabilizing a predetermined group of synchronous generators, referred to as targets. Since compromising each IBR requires time and resources, the adversary seeks to identify a subset of devices that can execute an attack most effectively. This motivates our limited-effort attack model, in which the adversary cannot compromise all IBRs and instead must select $m < h$ devices to gain control over. During an attack, the adversary manipulates the active power reference ($P_{ref,i}$ in \eqref{eq-s2ibrdroop}) of the $m$ chosen devices. This introduces an external control signal $u$ for each device that is represented by
\begin{equation} \label{eq-s2linsys}
    \dot{x} = \mathbf{A}x + \mathbf{B}u,
\end{equation}
where the columns of $\mathbf{B} \in \{0,1\}^{n\times m}$ are standard basis vectors corresponding to the $\Delta \omega_{H,i}$ states of the compromised devices and $u \in \mathbb{R}^m$ is the malicious reference signal that the adversary injects. The decision of which devices to compromise is represented by the matrix $\mathbf{B}$, and the formulated attack is represented by the signal $u$.
We consider a state-feedback law of the form
    \begin{equation*}
        u = -\mathbf{K}x,
    \end{equation*}
where the gain matrix $\mathbf{K} \in \mathbb{R}^{m \times n}$ defines the control law. 
Although this assumes that the malicious actor has perfect state information during the attack,~\cite{brown18} has shown that attacks under this formulation can be extended to only requiring local data using linear state observers~\cite{chen84}. We refer the reader to~\cite{brown18} for more details, and focus on state-feedback design for the remainder of this paper.

The attack model adopts the following assumptions:
\begin{enumerate}
    \item The linearized system analysis is valid until a machine is disconnected. Modeling the ensuing dynamics would require re-linearization around another steady state point.
    \item Feedback control does not saturate the power capability of any attacking IBRs. 
    \item The unstable eigenvalue $\tilde{\lambda}$, introduced in Section~\ref{Sub3b:opt}, is chosen separately from the attacking IBRs.
\end{enumerate}

\subsection{Attacker Selection Problem} \label{Sub3b:opt}

With the state feedback controller, the closed-loop system is given by $\mathbf{A}-\mathbf{B}\mathbf{K}$. Any closed-loop eigenvalue and eigenvector pair $(\lambda_i,v_i)$ satisfies
    \begin{equation*}
        (\mathbf{A}-\mathbf{B}\mathbf{K})v_i = \lambda_i v_i,
    \end{equation*}
or equivalently, by defining $w_i = \mathbf{K}v_i$, 
    \begin{equation} \label{eq-s3hautus-vw}
            (\mathbf{A}-\lambda_i \mathbf{I})v_i + \mathbf{B}w_i=0.
    \end{equation}

To execute a successful attack, an unstable eigenvalue $\tilde{\lambda} \in \mathbb{C}$ is placed such that the target states have large components in the eigenvector $\tilde{v} \in \mathbb{C}^n$ relative to the states of the compromised IBRs. This type of attack was introduced in~\cite{demarco96,demarco98}, and results in a greater amplitude in the targets' unstable behavior, causing them to be disconnected by protection devices while the compromised IBRs remain connected. This idea uses eigenvector assignment to shape the system's transient behavior~\cite{klein77}.

For any selected pair $(\tilde{\lambda},\tilde{v})$ that satisfies \eqref{eq-s3hautus-vw}, a corresponding gain matrix can be found. For a matrix $\mathbf{V}$ whose columns are $n$ independent eigenvectors $\mathbf{V}  =\begin{bmatrix} v_1 & \dots & v_n \end{bmatrix}$, if we compute the matrix $\mathbf{W} =\begin{bmatrix} w_1 & \dots & w_n \end{bmatrix}$ such that \eqref{eq-s3hautus-vw} is satisfied for all $i\in\{1,\dots,n\}$, then the gain matrix can be found as $\mathbf{K} = \mathbf{W}\mathbf{V}^{-1}$~\cite{subrahmanyan99}.

Let the set $\mathcal{T}$ contain the indices of the $\omega$ states of the target generators. The goal of the attacker is to maximize the objective function $z$ defined by the ratio
    \begin{equation*}
        z = \frac{\sum_{i\in \mathcal{T}}v^H_i v_i^{\vphantom{H}}}{v^H\mathbf{B}\mathbf{B}^\top v},
    \end{equation*}
where the notation $v^H$ represents the conjugate transpose of~$v$. The numerator represents the participation of the target states in the unstable mode, and the denominator represents the participation of the attacker (controlled) states. A large ratio indicates that the target states have a greater amplitude in the unstable mode than the attacker and will experience more severe oscillations. In other words, maximizing $z$ maximizes the unstable participation of the target states relative to the attacker states. In this ``successful attack'' scenario, the target devices will be tripped offline while the attacking devices remain online and may execute subsequent attacks. In subsequent formations, we show the equivalent minimization of the inverse objective~$z^{-1}$ to put the binary variable $\mathbf{B}$ in the numerator. 

The eigenvalue $\tilde{\lambda}$ determines the amplitude and frequency of the unstable oscillations generated by the attack and influences which targets are susceptible. Here, we assume that the complex eigenvalue $\tilde{\lambda}$ is a pre-selected constant parameter in the problem. The selection of $\tilde{\lambda}$ is discussed further in Section~\ref{Sub5d:selection}.

For any fixed $\tilde{\lambda}$, the corresponding eigenvector $\tilde{v}$ cannot be arbitrarily set and must satisfy \eqref{eq-s3hautus-vw}. Thus, the optimal attacker selection problem is given by
\begin{subequations} \label{eq-s3opt1}
    \begin{align}
        \min_{v,\,w,\,\mathbf{B}_{\rm ibr}}  \ \ &z^{-1} = \frac{v^H\mathbf{B}\mathbf{B}^\top v}{\sum_{i\in \mathcal{T}}v^H_i v_i^{\vphantom{H}}} \label{eq-s3opt1-a}\\ 
        \text{s.t.} \quad & (\mathbf{A}-\tilde{\lambda} \mathbf{I})v + \mathbf{B}w=0, \label{eq-s3opt1-b}\\
        & \mathbf{B}^\top \mathbf{1}_n  = \mathbf{1}_m, \label{eq-s3opt1-c}\\
        & \mathbf{B} = \begin{bmatrix}
            \mathbf{0}_{(n-h)\times m} \\ \mathbf{B}_{\rm ibr}
        \end{bmatrix}, \quad \mathbf{B}_{\rm ibr} \in \{0,1\}^{h\times m}, \label{eq-s3opt1-d} \\
        & v\in \mathbb{C}^n,\; w \in \mathbb{C}^m, \label{eq-s3opt1-e}
    \end{align}      
\end{subequations}
where $\mathbf{1}_n$ is the $n$-dimensional vector of ones. Equations~\eqref{eq-s3opt1-c} and \eqref{eq-s3opt1-d} define the structure of the matrix $\mathbf{B}$, whose $m$ columns are constrained to the standard basis vectors of the final $h$ states.  In previous works, the matrix $\mathbf{B}$ is assumed to be known, i.e., the attacking devices are specified rather than being a decision variable. References~\cite{demarco98} and~\cite{roberts21} have shown that for any fixed $\mathbf{B}$, \eqref{eq-s3opt1} resolves to a generalized eigenvalue problem with a closed-form solution. We will notate this solution as 
    \begin{equation} \label{eq-s3opt1-sol}
        z^* = f(\mathbf{B}),
    \end{equation}
where the function $f$ returns the inverse of the optimal objective of the minimization problem described by \eqref{eq-s3opt1} with a fixed $\mathbf{B}$ matrix. When $\mathbf{B}$ is a decision variable, as in the attacker selection problem we propose in this paper, the cross-terms between binary variables in $\mathbf{B}$ and complex continuous variables~$v$ significantly increase the problem's complexity. 

The first candidate solution approach is to iterate through all possible combinations of $m$ devices, computing the optimal objective for each using \eqref{eq-s3opt1-sol}, and selecting the best performing set. However, this brute-force search scales poorly when $h$ and $m$ are large, as there are $\binom{h}{m}$ total combinations. 

\subsection{MIQCP Attacker Selection} \label{Sub3c:MIQCP}

The optimization problem \eqref{eq-s3opt1} is nonlinear with a rational objective and cross-terms in the equality constraint \eqref{eq-s3opt1-b}. These complexities render it prohibitive for even state-of-the-art mixed-integer solvers. However, we can reformulate \eqref{eq-s3opt1} into an equivalent MIQCP which solvers such as Gurobi allow.

Instead of treating the matrix $\mathbf{B}$ as a decision variable, let $x \in \{0,1\}^n$ represent which states are compromised, where $x_i$ indicates whether the $i$-th state is maliciously controlled.
The number of attacking devices must equal $m$, and only the frequency states of the IBRs can be selected. Thus, we have
    \begin{align}
        &\mathbf{1}_n^\top x = m, \\
        & x = \begin{bmatrix}
            \mathbf{0}_{n-h}^\top & x_{\rm ibr}^\top
        \end{bmatrix}^\top.
    \end{align}
An interpretation of constraint \eqref{eq-s3opt1-b} is that the matrix-vector product $(\mathbf{A}-\tilde{\lambda} \mathbf{I})v$ lies in the subspace spanned by the columns of $\mathbf{B}$. Since $\mathbf{B}$ is only comprised of standard basis vectors corresponding to the attacker states, the subspace constraint can be written as the equivalent relation
    \begin{equation} \label{eq-s3opt2sub}
        (\mathbf{I}-\mathbf{X})(\mathbf{A}-\tilde{\lambda} \mathbf{I})v=0,
    \end{equation}
with the notation $\mathbf{X} = \mathsf{diag}(x)$, which states that all unselected entries of $(\mathbf{A}-\tilde{\lambda} \mathbf{I})v$ must be zero.
The objective function is scale-invariant in $v$, and the denominator can be removed by fixing its magnitude to 1. Furthermore, the matrix $\mathbf{B}$ can be replaced by observing that $\mathbf{B}\mathbf{B}^\top = \mathbf{X}$. This intermediate reformulation with the new variable $x$ is expressed as
    \begin{align}\label{eq-s3opt2}
        \min_{v,\,x_{\rm ibr}} \ \ & v^H\mathbf{X} v \nonumber\\ 
        \text{s.t.} \quad & \sum_{i\in \mathcal{T}}v_i^Hv_i^{\vphantom{H}} = 1, \nonumber\\
        & (\mathbf{I}-\mathbf{X})(\mathbf{A}-\tilde{\lambda} \mathbf{I})v=0, \nonumber\\
        & \mathbf{1}_n^\top x = m, \quad x = \begin{bmatrix}
            \mathbf{0}_{n-h}^\top & x_{\rm ibr}^\top
        \end{bmatrix}^\top, \nonumber\\
        & x_{\rm ibr} \in \{0,1\}^{h}, \;v \in \mathbb{C}^n.
    \end{align}
Each component of the eigenvector contributes to the new objective function its squared magnitude if the component corresponds to a compromised state, and zero otherwise. This relation can instead be written as a logical constraint. By introducing $c_i = x_i^{\vphantom{H}} v_i^H v_i^{\vphantom{H}}$, the objective becomes $\sum_i^n c_i$, and the constraints
\begin{equation*}
c_i \geq 0, \ \ \ c_i \geq v_i^Hv_i^{\vphantom{H}} - M(1-x_i),
\end{equation*}
are added for all $i \in \{1,\dots,n\}$, where $M$ is a large (big-$M$) constant. The binary-continuous products are thus replaced by a linear objective and mixed-integer convex inequalities. Similarly, the subspace constraints \eqref{eq-s3opt2sub} can be rewritten as linear big-$M$ constraints as
\begin{equation}
        -M(1-x) \leq (\mathbf{A}-\tilde{\lambda} \mathbf{I})v \leq M(1-x).
\end{equation}
With this, the problem in \eqref{eq-s3opt2} is transformed into the MIQCP
\begin{subequations} \label{eq-s3opt3-MIQCP}
\begin{align} 
    \min_{v,\,x_{\rm ibr},\,c} \ \ & \sum_{i=1}^n c_i \label{eq-s3opt3-a}\\ 
    \text{s.t.} \quad & \sum_{i\in \mathcal{T}}v_i^Hv_i^{\vphantom{H}} = 1, \label{eq-s3opt3-b}\\
    & -M(1-x) \leq (\mathbf{A}-\tilde{\lambda} \mathbf{I})v \leq M(1-x), \\
    & \mathbf{1}_n^\top x = m,\quad x = \begin{bmatrix}
            \mathbf{0}_{n-h}^\top & x_{\rm ibr}^\top
        \end{bmatrix}^\top, \\
    & c_i \geq v_i^Hv_i - M(1-x_i), \\
    & c_i \geq 0, \qquad \qquad \qquad \qquad \qquad \forall i \in \{1,\dots,n\}\\
    & x_{\rm ibr} \in \{0,1\}^{h}, \; v \in \mathbb{C}^n, \; c \in \mathbb{R}^n,
\end{align}    
\end{subequations}
with the normalization condition \eqref{eq-s3opt3-b} remaining the sole nonconvex constraint. When only one machine is designated as the target, \eqref{eq-s3opt3-b} is resolved and the problem is simplified to a mixed-integer convex program. We elaborate on this special but nontrivial case and provide a discussion of target selection in Section~\ref{Sub5d:selection}.

The values of the big-$M$ constants must be set somewhat arbitrarily, as the values of $v$ are otherwise unbounded.
However, the normalization constraint on the target states of $v$ informs reasonable values of $M$, since the attackers would like those states to have the largest magnitudes in the vector.

For some systems, the problem in \eqref{eq-s3opt3-MIQCP} can be directly solved by state-of-the-art solvers such as Gurobi~\cite{gurobi23}. 
However, for larger systems with many candidate compromisable IBRs, both the combinatorial optimization problem~\eqref{eq-s3opt3-MIQCP} and brute-force searching may not be computationally practical. In the next section, we present a heuristic selection approach based on the special case of selecting a single attacker.

\section{Attack Capability Ranking Heuristic}\label{Sec4:heuristic}

This section discusses a special case in which only one attacking device is selected from the system's $h$ IBRs, that is, the attack budget $m=1$. Section~\ref{Sub4a:single} shows how the attack selection problem in this special case simplifies to comparing the ratios of matrix diagonal terms. Building on this, Section~\ref{Sub4b:heuristic} develops a heuristic algorithm for the general problem based on the attacking capability of each IBR when working independently.

\subsection{Single Attacker Scenario} \label{Sub4a:single}

In the initial formulation \eqref{eq-s3opt1} when $m=1$, the input matrix $\mathbf{B}$ is an $n$-dimensional vector that we will notate as $\mathbf{b} \in \{0,1\}^n$, and $w$ is a complex scalar. Thus, \eqref{eq-s3opt1} reduces to
    \begin{subequations} \label{eq-s4opt4}
        \begin{align}
            \min_{v,\, w,\, \mathbf{b}_{\rm ibr}}  \ \ & z^{-1} =\frac{v^H\mathbf{b}\mathbf{b}^\top v}{\sum_{i\in \mathcal{T}}v^H_i v_i^{\vphantom{H}}} \label{eq-s4opt4-a}\\ 
            \text{s.t.} \quad & (\mathbf{A}-\tilde{\lambda} \mathbf{I})v + \mathbf{b}w=0, \label{eq-s4opt4-b}\\
            & \mathbf{b}^\top \mathbf{1}_n  = 1, \label{eq-s4opt4-c}\\
            & \mathbf{b} = \begin{bmatrix}
            \mathbf{0}_{(n-h)}^\top & \mathbf{b}_{\rm ibr}^\top
            \end{bmatrix}^\top, \quad \mathbf{b}_{\rm ibr} \in \{0,1\}^{h}, \label{eq-s4opt4-d} \\
            & v\in \mathbb{C}^n, \; w \in \mathbb{C}. \label{eq-s4opt1-e}
        \end{align}      
    \end{subequations}
For simplicity of notation, let $\mathbf{\hat{A}}=(\mathbf{A} - \tilde{\lambda} \mathbf{I})^{-1}$, and define the matrix $\mathbf{C} = [\hat{\mathbf{e}}_i]_{i \in \mathcal{T}}$ that selects the target states of $v$. The denominator in~\eqref{eq-s4opt4-a} can be rewritten as $v^H\mathbf{C}\mathbf{C}^\top v$. 
By substituting $v$ in the objective using \eqref{eq-s4opt4-b}, both $v$ and $w$ can be eliminated from the problem, leaving only binary variables. Furthermore, constraints \eqref{eq-s4opt4-c} and \eqref{eq-s4opt4-d} restrict $\mathbf{b}$ to a subset of standard basis vectors. The problem thus simplifies to
    \begin{subequations} \label{eq-s4opt5}
    \begin{align}
        \min_{b} \ \ &\frac{(\mathbf{b}^\top\mathbf{\hat{A}}^H\mathbf{b})(\mathbf{b}^\top\mathbf{\hat{A}}\mathbf{b})}{\mathbf{b}^\top\mathbf{\hat{A}}^H\mathbf{C}\mathbf{C}^\top\mathbf{\hat{A}}\mathbf{b}}  \label{eq-s4opt5-a}\\ 
        \text{s.t.} \quad & \mathbf{b} \in \{\hat{\mathbf{e}}_{n-h+1},\dots,\hat{\mathbf{e}}_n\}.\label{eq-s4opt5-b}
    \end{align}
    \end{subequations}
The product $\hat{\mathbf{e}}_i^\top\mathbf{X}\hat{\mathbf{e}}_i$ extracts the $i$-th diagonal value of $\mathbf{X}$. Thus, the optimal ratio $z^*$ and reciprocal of the solution to \eqref{eq-s4opt5} is given by the unconstrained maximization problem
    \begin{equation} \label{eq-s4opt6}
        z^* =\max_{i \in \{n-h+1,\dots,n\}} \ \ \frac{(\mathbf{\hat{A}}^H\mathbf{C}\mathbf{C}^\top\mathbf{\hat{A}})_{ii}}{\mathbf{\hat{A}}^H_{ii}\mathbf{\hat{A}}_{ii}},
    \end{equation}
which can be solved efficiently by comparing $h$  ratios of matrix diagonal terms. The ratio $z_i = \frac{(\mathbf{\hat{A}}^H\mathbf{C}\mathbf{C}^\top\mathbf{\hat{A}})_{ii}}{\mathbf{\hat{A}}^H_{ii}\mathbf{\hat{A}}_{ii}}$ is a metric of how effectively IBR $i$ can carry out a malicious attack alone.

\subsection{The Row-Search Heuristic}  \label{Sub4b:heuristic}

We return to the general case of multiple compromised devices. Extending the single-attacker case, one natural strategy may be to control the $m$ attackers that perform best individually. We term this the "single best" heuristic, with an attacking set $\mathcal{C}_1$ that contains the indices of the $m$ best devices as
\begin{equation} \label{eq-s4heur1set}
    \mathcal{C}_1 =\{i:z_i\in\mathrm{max}_m(z_1,\dots,z_h)\}.
\end{equation}
The corresponding input matrix is formulated as $\mathbf{B} = [\hat{\mathbf{e}}_i]_{i\in \mathcal{C}_1}$.

In practice, this heuristic may perform sub-optimally, as the most effective IBRs at attacking alone may not be the optimal choices when attacking as a group. The interactions between the attacking devices are not captured by \eqref{eq-s4opt6}, which, when considering multiple attackers, includes expressions of the off-diagonal terms of the matrices. 
Alternatively, $\mathcal{C}_1$ can be used as an initial candidate set to be refined using a partial search, as shown in Algorithm~\ref{alg-heur-row}.

\begin{algorithm}
\caption{Row Search Heuristic}\label{alg-heur-row}
\begin{algorithmic}[1]
\State Given: $\mathbf{A}$, unstable eigenvalue $\tilde{\lambda}$, and target matrix $\mathbf{C}$
\State Compute $\mathbf{\hat{A}}=(\mathbf{A}-\tilde{\lambda}\mathbf{I})^{-1}$
\State Initialize $\mathcal{C}_{alg} =\mathcal{C}_1$ from \eqref{eq-s4heur1set}, and formulate $\mathbf{B}$
\State Initialize $z_{\rm best} = f(\mathbf{B})$
\State swapped = TRUE
\While{swapped = TRUE}
\State swapped = FALSE
\For{$j\in\mathcal{C}_{alg}$}
\For{$k=\{n-h+1,\dots,n\}\setminus\mathcal{C}_{alg}$}
\State $\mathcal{C}= k \cup (\mathcal{C}_{alg}\setminus j)$
\State Solve \eqref{eq-s3opt1-sol} for~$z^*$ using $\mathbf{B} = [\hat{\mathbf{e}}_i]_{i\in \mathcal{C}}$
\If{$z^* > z_{\rm best}$}
\State Set $\mathcal{C}_{\rm best} = \mathcal{C}$, $z_{\rm best} = z^*$
\State swapped = TRUE
\EndIf
\EndFor
\EndFor
\State $\mathcal{C}_{alg} = \mathcal{C}_{\rm best}$, $z_{alg} = z_{\rm best}$
\EndWhile
\State Output the refined set $\mathcal{C}_{alg}$ and objective value $z_{alg}$
\end{algorithmic}
\end{algorithm}

Algorithm \ref{alg-heur-row} begins with the set of IBRs obtained from \eqref{eq-s4heur1set}, with an initial objective value of $f(\mathbf{B})$. The algorithm then considers swapping out any member of the current set with an unselected device. If any swap improves the objective value, the device is marked as a candidate. After all swaps have been considered, the best swap is executed, and the algorithm continues until no new swaps occur.

The algorithm is guaranteed to converge, as the objective value monotonically increases with each iteration. 
When $h$ and $m$ are large, Algorithm~\ref{alg-heur-row} provides an efficient alternative to brute-force searching and the MIQCP~\eqref{eq-s3opt3-MIQCP}. The search process evaluates a much smaller subset of combinations than the exhaustive brute-force approach. While the heuristic is not guaranteed to give an optimal attacking set, Section~\ref{Sec5:Case-Study} empirically shows that the algorithm performs exceptionally well while significantly reducing the computational burden.

\section{Case Study}\label{Sec5:Case-Study}

This section demonstrates the entire attacker selection and attack formulation process and illustrates its effect via nonlinear dynamical simulations. The main case study is conducted on the WSCC 179-bus system, with additional supporting results from the \mbox{ACTIVSg500} system. We show that the heuristic algorithms produce near-optimal attacking sets for a variety of scenarios while dramatically reducing the required computational time. Finally, we discuss how the selection of targets and unstable eigenvalue affect the attack's potency.

\subsection{Test Systems}
\begin{figure}[htbp]
        \centering
        \includegraphics[width=\linewidth]{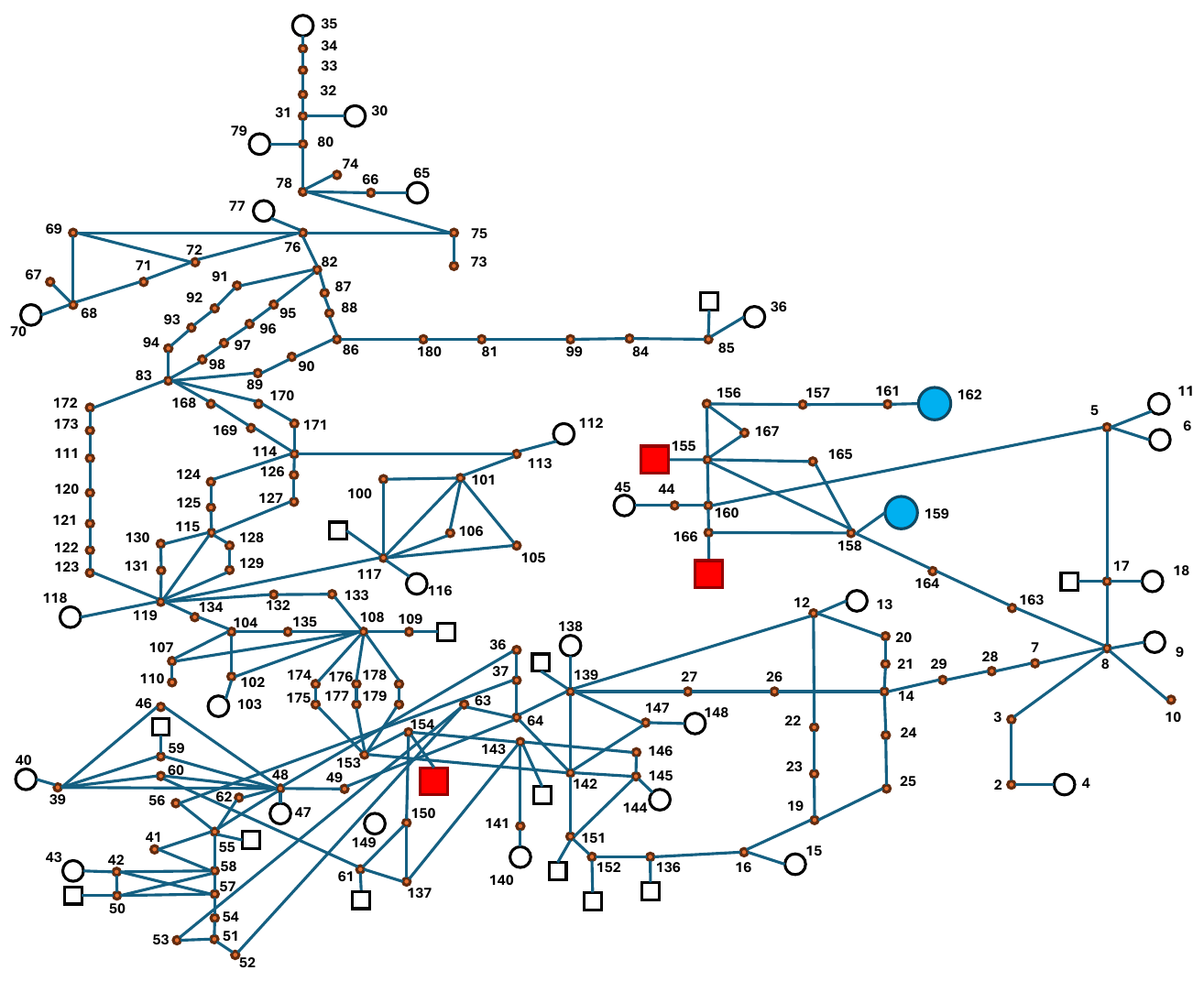}
        \caption{Modified WSCC 179-bus System. Blue: target generators, red: compromised IBRs in the example attack.}
        \label{fig-wscc179}
\end{figure}

The primary case study uses a modified version of the WSCC 179-bus system shown in Figure~\ref{fig-wscc179}. In addition to the original 29 synchronous generators, 16 IBRs are introduced at various buses in the system with the circuit model described in~\cite{du23}. The dynamic parameters of the generators are based on those of~\cite{brown18}, and the IBR droop coefficients and time constants are selected within reasonable ranges \cite{du23}. The parameters of all network components and the initial operating point are provided in~\cite{zou2026}.

The system is linearized around a stable operating point to obtain the state matrix $\mathbf{A}\in \mathbb{R}^{264\times 264}$. States 30-58 and 249-264 represent the frequency variables of the synchronous generators and IBRs, respectively. The matrix has 86 real eigenvalues and 89 complex eigenvalue pairs. The oscillatory modes between the generators range from 2.5 to 9.4~rad/s (0.4 to 1.5~Hz). 

We also utilize \mbox{ACTIVSg500}, a 500-bus synthetic system of South Carolina. The system contains 60 synchronous generators and is modified to include 30 vulnerable IBRs. We adopt the dynamic model~\eqref{eq-s2fluxdecay}--\eqref{eq-s2network} and set dynamic parameters to values within their typical ranges as provided in~\cite{zou2026}.

\subsection{Example Attack} \label{Sub5b:attack}
We illustrate the attack formulation process from start to finish on the WSCC 179-bus system. The target states are selected as $\mathcal{T}=\{57, 58\}$ and the unstable eigenvalue is chosen as $\tilde{\lambda} = 0.2 + j5.12$. We devise an attack in which $m=3$ IBRs can be compromised.

By solving the MIQCP in \eqref{eq-s3opt3-MIQCP}, the best attacking set is found as $\mathcal{C} = \{262,263,264\}$. The target and attacker devices are shown in green and blue in Figure~\ref{fig-wscc179}, respectively. The optimal objective of $z^* = 6.357$ indicates that the aggregated participation of the target machines is $z^*$ times greater than that of the attacking devices. This can also be seen in the optimal eigenvector, the target and attacker components of which are shown in Table~\ref{tab-eigenvector}.

\begin{table}[t]
    \centering
    \caption{Closed-loop eigenvector components of select states}
    \vspace{5mm}
    \begin{tabular}{|c | c | c |} 
    \hline
    State & $v$ & $|v|$\\ 
    \hline
    57 (t1) & $-0.3742 - j0.0671$ & $0.380$\\
    \hline
    58 (t2) &  $-0.5679 - j0.1093$ & $0.578$\\
    \hline
    262 (a1) & \ \ $0.0408 - j0.1715$ &     $0.176$ \\
    \hline
    263 (a2) & \ \ $0.0221 - j0.1479$ & 
    $0.150$ \\
    \hline
    264 (a3) & \ \   $0.0302 - j0.1448$ & $
    0.148$ \\
    \hline
    \multicolumn{3}{|c|}{$z^* = {\sum_{i\in\mathcal{T}}|v_i|^2} \div{\sum_{j\in\mathcal{C}}|v_j|^2} = 6.357$} \\
    \hline
    \end{tabular}
    \label{tab-eigenvector}
\end{table}

\begin{figure}[t]
        \centering
        \includegraphics[width=\linewidth]{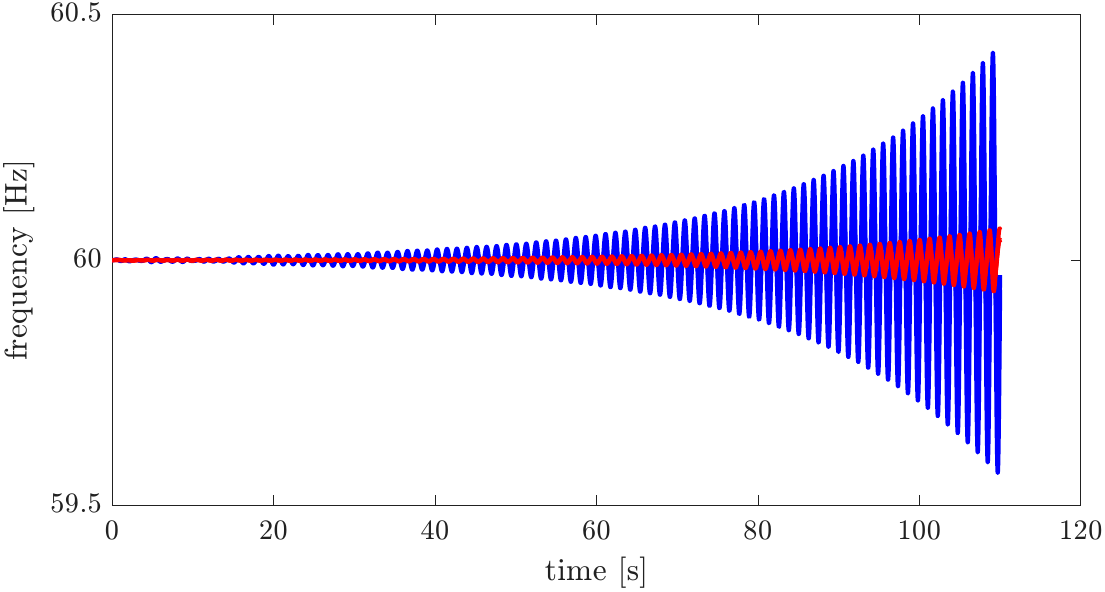}
        \caption{DAE simulation of the WSCC 179-bus system with optimally selected attackers. Blue: target frequency states. Red: attacker frequency states.}
        \label{fig-dae-optimal}
\end{figure}

The state feedback matrix $\mathbf{K} \in \mathbb{R}^{3\times264}$ is derived by replacing an existing mode with the unstable mode $(\tilde{\lambda} ,\tilde{v})$. To verify that the system is destabilized by the attack, the nonlinear DAE described by~\eqref{eq-s2fluxdecay}--\eqref{eq-s2network} in Section~\ref{Sec2:Prob} is simulated with malicious feedback control at the chosen IBRs. The system is subjected to a temporary 2-second line outage, and the attack begins when the line is restored. Figure~\ref{fig-dae-optimal} shows the frequency states of the target synchronous generators (blue) and the attacking IBRs (red). The simulation ends when the rate of change of a target generator's frequency (ROCOF) exceeds 2~Hz/s, the threshold adopted by \cite{brown18} at which protective devices will disconnect a synchronous generator. This value is supported by \cite{uijlings13}, which found that synchronous generators generally cannot maintain stability with a ROCOF greater than 2~Hz/s.

As expected, the electrical frequency states of both the attacking devices and target generators exhibit growing oscillations at a frequency around $\mathrm{Im}(\tilde{\lambda})=5.12$~rad/s. The amplitude of the targets' swings is much greater than that of the attackers. The ratio of summed squared amplitudes matches the value predicted by the optimal objective function $z^*=6.357$. The attack results in the disconnection of generator 29, which will subject the network to further dynamic transients. 
The linearized system may not accurately represent the system's dynamics following the removal of a generator, which alters the original nonlinear system. How the system evolves and how a continued attack may ensue thereafter is a subject of future work.

\begin{figure}[t]
        \centering
        \includegraphics[width=\linewidth]{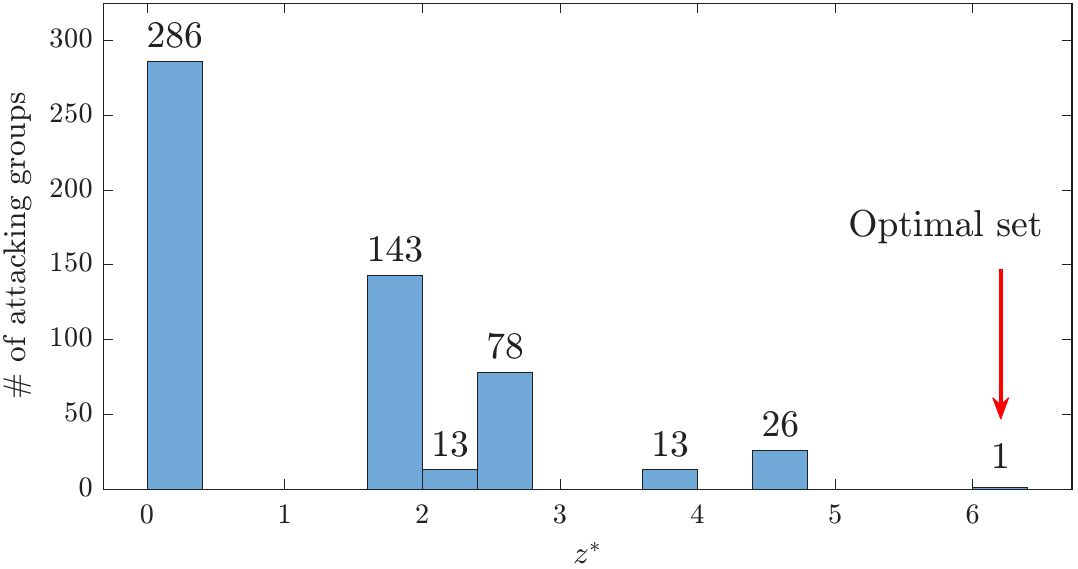}
        \caption{Histogram of the performance of all 3-member attacking sets. The optimal set with $z^* = 6.357$ performs much better than all others.}
        \label{fig-histogram_sets}
\end{figure}

To illustrate the importance of optimally selecting the attacking devices, the greatest attainable objective value for each set of $m=3$ attackers is plotted in a histogram in Figure~\ref{fig-histogram_sets}. 

As this figure shows, the optimal set, whose attack is illustrated in Figure~\ref{fig-dae-optimal}, out-performs all other three-device combinations by a wide margin.
A majority of attacking groups can only achieve a ratio less than 1, which indicates that affecting the target synchronous generators requires these groups to inflict a more severe instability onto themselves, rendering such attacks ineffective. Thus, these results highlight that not all groups of compromised IBRs can execute a successful attack for a set of designated targets. The choice of which IBRs to compromise directly determines the level of threat an adversary may pose to the system.

\subsection{Evaluation of Attacker Selection Heuristics}

Three metrics are used to evaluate the effectiveness of the heuristic algorithms:
\begin{enumerate}
    \item Does the heuristic identify the optimal control set?
    \item How well does the heuristic attacking set perform compared to the optimal one?
    \item Does the heuristic reduce the computational time of the problem?
\end{enumerate}

To show the effectiveness of the heuristic selection algorithms from Section~\ref{Sec4:heuristic}, we solve the problem for every combination of two and three synchronous generators as targets for the WSCC 179-bus system, and two synchronous generators as targets for the \mbox{ACTIVSg500} system. In each case, a suitable eigenvalue $\tilde{\lambda}$ is first selected, and the optimization problem~\eqref{eq-s3opt3-MIQCP} and Algorithm~\ref{alg-heur-row} are used to find the best group of $m=3$ IBRs as attackers. The aggregated results are shown in Table~\ref{tab-heuristics}, along with the brute force approach, which searches through all possible solutions.

\begin{table}[ht]
    \centering
    \caption{Performance metrics of heuristic algorithms}
    \begin{tabular}{|c | c | c | c |} 
    \hline
    Algorithm & Find optimal set & $z^*/z^*_{\rm optimal}$ & Avg. time (s) \\
    \hline
    \multicolumn{4}{|c|}{WSCC-179: 2 targets (406 scenarios)} \\
    \hline
    Brute search & $100\%$ & $100\%$ & $0.0120$\\
    \hline
    Single best \eqref{eq-s4heur1set} & $87.44\%$ & $99.59\%$ & $0.0073$\\
    \hline
    Row-search (Alg. \ref{alg-heur-row}) &  $99.75\%$ & $\sim100\%$ & $0.0092$\\
    \hline
    MIQCP (Eq. \eqref{eq-s3opt3-MIQCP}) &  $100\%$ & $100\%$ & $23.044$\\
    \hline
    \multicolumn{4}{|c|}{WSCC-179: 3 targets (3654 scenarios)} \\
    \hline
    Brute search & $100\%$ & $100\%$ & $0.0141$\\    \hline
    Single best \eqref{eq-s4heur1set} & $82.40\%$ & $99.43\%$ & $0.0090$\\
    \hline
    Row-search (Alg. \ref{alg-heur-row}) &  $99.78\%$ & $\sim100\%$ & $0.0099$\\
    \hline
    MIQCP (Eq. \eqref{eq-s3opt3-MIQCP}) &  $100\%$ & $100\%$ & $43.962$\\
    \hline
    \multicolumn{4}{|c|}{ACTIVSg500: 2 targets (1770 scenarios)} \\
    \hline
    Brute search & $100\%$ & $100\%$ & $0.0637$\\    \hline
    Single best \eqref{eq-s4heur1set} & $98.98\%$ & $99.99\%$ & $0.0401$\\
    \hline
    Row-search (Alg. \ref{alg-heur-row}) &  $100\%$ & $100\%$ & $0.0415$\\
    \hline
    MIQCP \eqref{eq-s3opt3-MIQCP} &  $100\%$ & $100\%$ & $112.526$\\
    \hline
    \end{tabular}
    \label{tab-heuristics}
\end{table}

\begin{figure}[h]
        \centering
        \includegraphics[width=\linewidth]{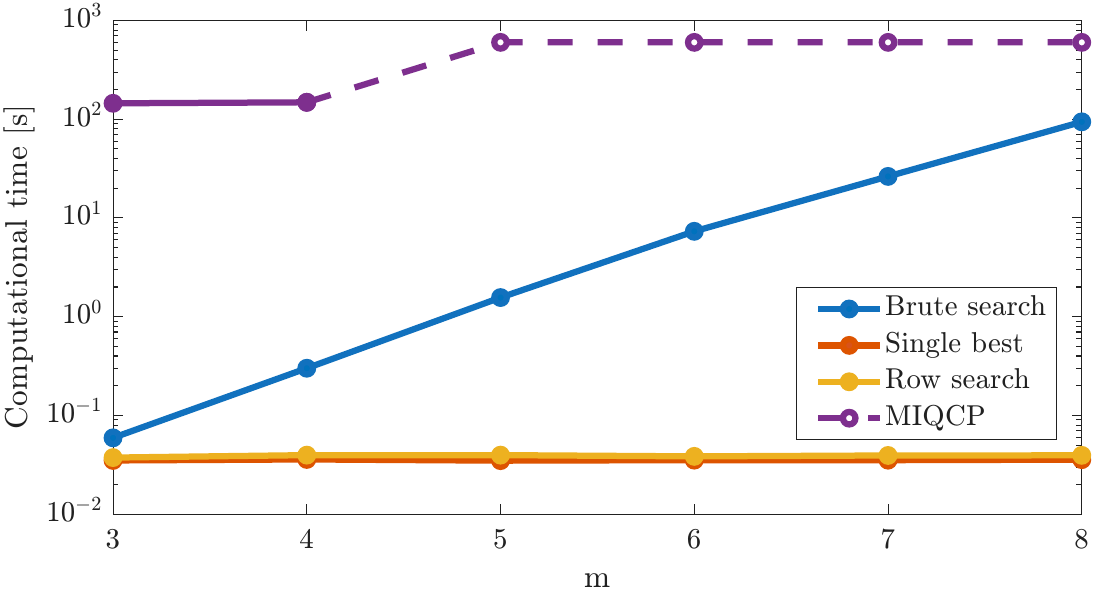}
        \caption{A 10-minute limit was enforced for the MIQCP, with unfilled markers indicating time limit was reached. The optimal methods scale poorly while the heuristics use around the same amount of time.}
        \label{fig-increase_m}
\end{figure}

As Table~\ref{tab-heuristics} shows, the single best heuristic is able to identify the optimal attacking set in a majority of the scenarios. On average, its solutions perform well compared to the optimal solution, even when it does not find the optimal set. The row search heuristic improves upon this, obtaining the optimal set in almost every single scenario at a small additional computational cost. Both heuristics reduce the computational time compared to the optimal methods, a reduction that becomes greater for larger values of $m$. Figure~\ref{fig-increase_m} shows this trend, comparing the average computational times of 50 2-target scenarios from the ACTIVSg500 system. 

For greater values of $m$, the brute search method slows down significantly, having to check more combinations. In contrast, the heuristic algorithms require about the same amount of time to complete as they consider only a small subset of combinations. Furthermore, the heuristics continue to identify near-optimal solutions akin to the results in Table~\ref{tab-heuristics}, showing the effectiveness of the proposed heuristics.

The MIQCP approach solves the problem optimally, but is much slower than even brute search. This may be caused by the higher dimensionality of the formulation. The brute force method solves \eqref{eq-s3opt1-sol} for all permissible values of the input matrix~$\mathbf{B}$. However, the generalized eigenvalue problem described by \eqref{eq-s3opt1-sol} is $m$-dimensional. In contrast, the MIQCP in \eqref{eq-s3opt3-MIQCP} contains $n$-dimensional variables and cannot be simplified to a lower dimension. Thus, when $n \gg m$, the MIQCP is outperformed by brute search. However, we present the MIQCP as a basis for future work which is elaborated upon in Section~\ref{Sec6:Conclusion}.

\subsection{Target and Eigenvalue Selection} \label{Sub5d:selection}

This section examines how the selection of the targets and the unstable eigenvalue affect attack severity.
We use examples from the WSCC 179-bus system to show how the unstable eigenvalue affects the attack's severity and how each machine in the target set is affected.
We present an intuition-based method to select the eigenvalue and leave a more rigorous analysis as a subject of future work.

The real and imaginary components of the eigenvalue dictate the exponential rate of growth and frequency of the injected unstable mode, respectively. To set the value of real component, the adversary must consider how quickly the instability must grow to disconnect the target(s). The longer an attack continues, the more likely it is to be detected. The choice of the imaginary component plays a major role in the attack's effectiveness. Numerical simulations show that the best value lies close to the frequency of an existing oscillatory mode. Therefore, these frequencies provide candidate values for the eigenvalue's imaginary component. As a heuristic, we propose solving the attacker selection problem with each candidate value to identify the best choice for the eigenvalue's imaginary component.

\begin{figure}[t]
        \centering
        \includegraphics[width=\linewidth]{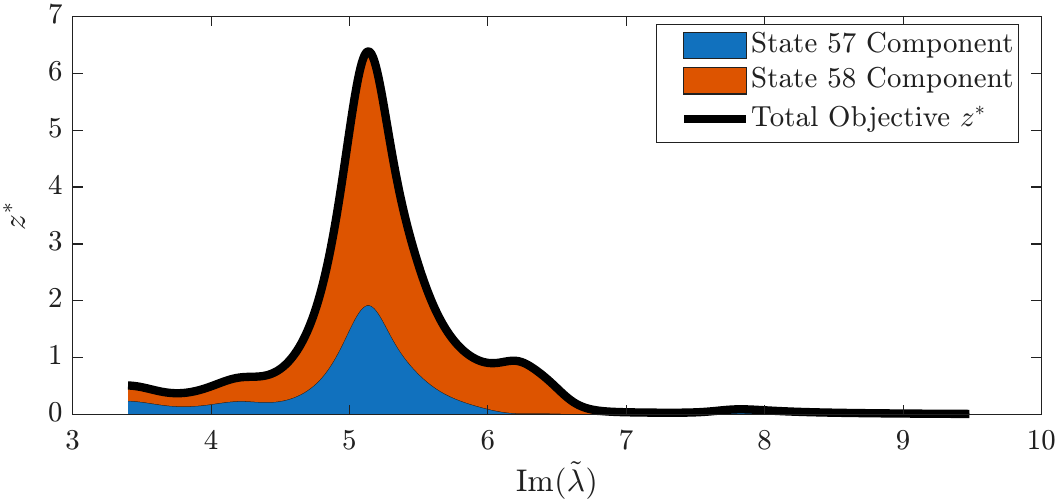}
        \caption{Optimal objective function for the case study in \ref{Sub5b:attack} with various values of the placed eigenvalue's imaginary part $\mathrm{Im}(\tilde{\lambda})$.}
        \label{fig-example1_lambda}
\end{figure}

From the case study in Section~\ref{Sub5b:attack} with the target set $\mathcal{T} = \{57,58\}$, the eigenvalue of the attack was selected as $\tilde{\lambda} = 0.2+j5.12$. Figure~\ref{fig-example1_lambda} shows the optimal objective of \eqref{eq-s3opt1} for different choices of $\mathrm{Im}(\tilde{\lambda})$ with a fixed $\mathrm{Re}(\tilde{\lambda}) = 0.2$.
The black curve shows the variation of the optimal objective function, and the colors underneath show how each target state contributes to it.
The optimal objective changes as the imaginary component varies. The peak occurs at $j5.12$, the value from the case study, at which both target states contribute substantially to the objective $z^*$. The participation factors of the original system show that both target machines participate in a mode with frequency $j5.14$. This indicates that generators are more susceptible to attacks close to their natural modes of oscillation, but not necessarily their most participated mode.

\begin{figure}[t]
        \centering
        \includegraphics[width=\linewidth]{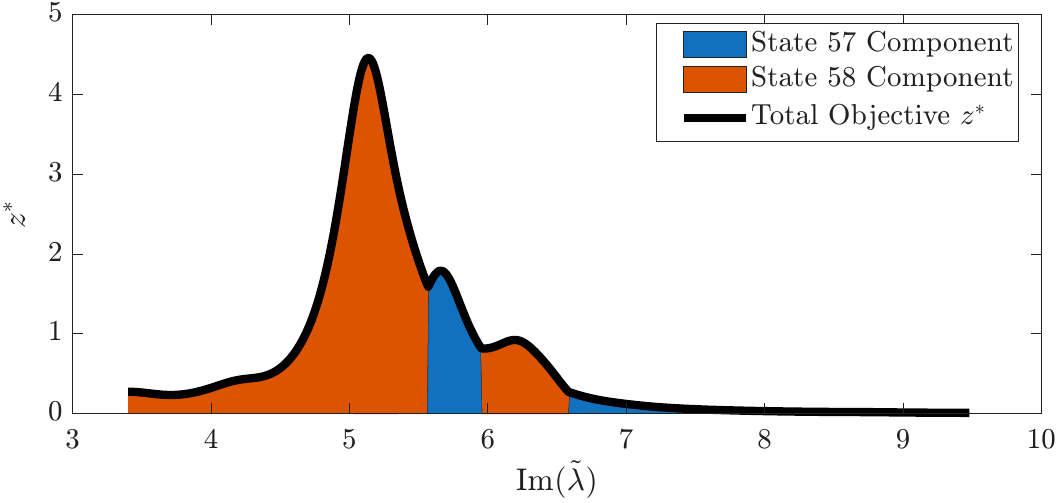}
        \caption{Optimal objective function for the target set $\mathcal{T}=\{53,58\}$ with various values of the placed eigenvalue $\tilde{\lambda}$.}
        \label{fig-example2_lambda}
\end{figure}

In the next example, the target set is altered to be $\mathcal{T}=\{53,58\}$, and the optimal objective is plotted again for various $\mathrm{Im}(\tilde{\lambda})$ with a fixed $\mathrm{Re}(\tilde{\lambda})=0.2$ in Figure~\ref{fig-example2_lambda}.
The objective function peaks at $j5.14$, but despite the high overall objective value, only state 58 is significantly affected by the attack. A smaller peak at $j5.69$ affects only state 53. This shows that although the optimization framework in \eqref{eq-s3opt1} maximizes the aggregated magnitudes of $v$ in the target states, it is agnostic of how each target is affected. When multiple targets are designated, the formulated attack may affect all of them, as in Figure~\ref{fig-example1_lambda}, or focus on a subset as in Figure~\ref{fig-example2_lambda}.

\begin{figure}[t]
        \centering
        \includegraphics[width=\linewidth]{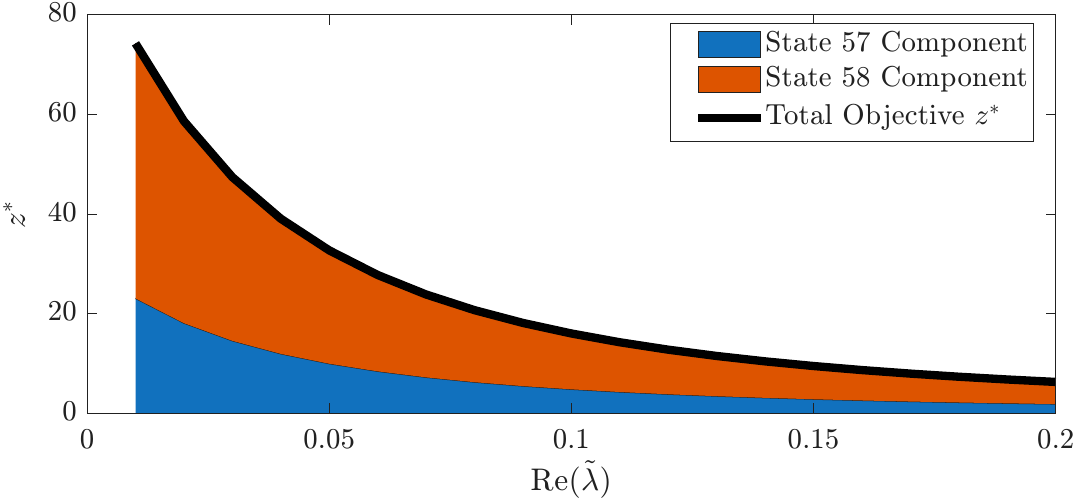}
        \caption{Optimal objective function for the case study in \ref{Sub5b:attack} with various values of the placed eigenvalue's real part $\mathrm{Re}(\tilde{\lambda})$.}
        \label{fig-real_lambda}
\end{figure}

The eigenvalue's real part controls the unstable oscillation's growth rate. To show its effect, the problem is computed with the target set $\mathcal{T}=\{57,58\}$, with the eigenvalue's imaginary component fixed at $\mathrm{Im}(\tilde{\lambda}) = j5.12$ in Figure~\ref{fig-real_lambda}. 
As the real part of $\tilde{\lambda}$ is increased, the optimal objective decreases, indicating a trade-off between the rate of unstable growth and the participation of the targets relative to the attackers. The malicious actor can choose this value to balance the two factors to influence how quickly the attack unfolds.

\begin{figure}[t]
        \centering
        \includegraphics[width=\linewidth]{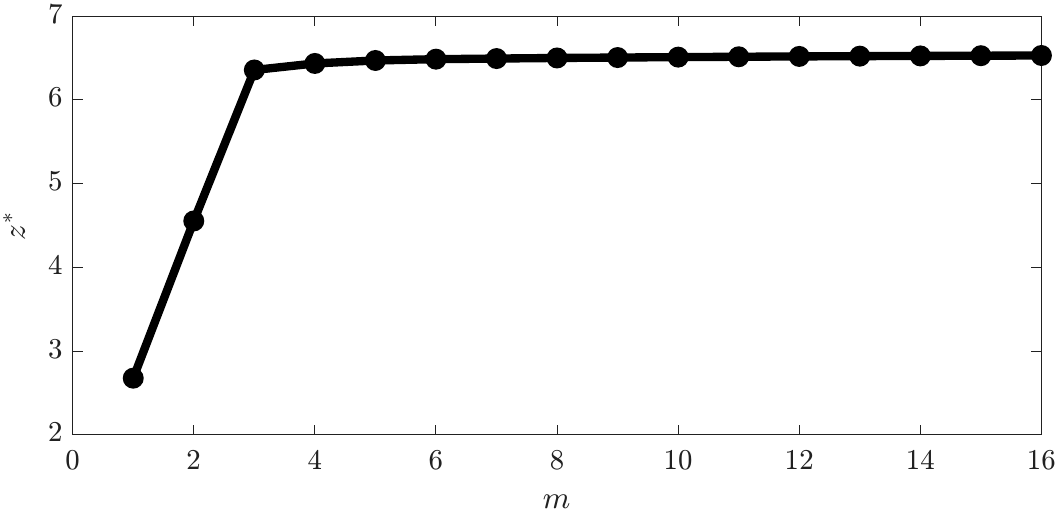}
        \caption{Optimal objective function for the case study in Section~\ref{Sub5b:attack} with various sizes of the attacking (controlled) set.}
        \label{fig-m}
\end{figure}

Finally, Figure~\ref{fig-m} shows the effect of increasing the number of devices $m$ that can be compromised.
Although compromising more IBRs increases the potency of an attack, there is diminishing value in every additional device compromised. In this case, the adversary sees negligible improvement when compromising more than three IBRs. This further reinforces our limited effort model, indicating that the selection of compromised devices is a much greater factor to a successful attack than their number.

\section{Conclusion}\label{Sec6:Conclusion}

This paper investigated how different sets of compromised IBRs can destabilize a prescribed set of target generators. This is cast as a combinatorial optimization program that selects both which devices to compromise and how they should be maliciously controlled. An alternative MIQCP formulation and a row-search heuristic are also presented. The adopted solution approach depends on the problem's scale and desired optimality guarantees. 
The attacker selection process is demonstrated in detail using the WSCC 179-bus network, with supplementary results for the \mbox{ACTIVSg500} system. The results show that many combinations of compromised devices are unable to successfully execute an attack, highlighting the importance for an adversary to select the most potent group. We further show how the problem is influenced by the selection of the unstable eigenvalue, target set, and number of compromised IBRs.
By understanding which devices may contribute to a cyber-physical attack, the goal of this work is to uncover the vulnerabilities of a system and inform where cybersecurity protection systems are most crucial.

Future work may investigate relaxation or decomposition approaches to resolve the nonconvex normalization constraint \eqref{eq-s3opt3-b} that remains in the MIQCP formulation~\eqref{eq-s3opt3-MIQCP}. Although this formulation can be solved by commercial solvers, eliminating the nonconvex constraint would greatly simplify the problem and improve its computational time.

Another future direction is to explore the system dynamics after a synchronous generator is disconnected. Our analysis relies on a linearization around an operating point, an approximation that may break down when the system changes substantially, like when one or more generators are removed. Increasing scope to consider system behavior after an attack successfully causes generator disconnections would model an even more sophisticated attacker than considered in this paper.

\bibliographystyle{ieeetr}
\bibliography{refs}

@techreport{irena_digitalisation_ai_2025,
  title       = {Digitalisation and {AI} for Power System Transformation:
                 Perspectives for the {G7}},
  institution = {{International Renewable Energy Agency}},
  year        = {2025},
  note        ={\url{https://www.irena.org/-/media/Files/IRENA/Agency/Publication/2025/Oct/IRENA_INN_Digitalisation_AI_for_power-systems_2025.pdf}}
}

@techreport{irena_capacity_highlights_2025,
  title       = {Renewable Capacity Highlights 2025},
  institution = {International Renewable Energy Agency},
  year        = {2025},
  note        ={\url{https://www.irena.org/-/media/Files/IRENA/Agency/Publication/2025/Mar/IRENA_DAT_RE_Capacity_Highlights_2025.pdf}}
}

@techreport{iea_renewables_2024,
  title       = {Renewables 2024: {Analysis} and Forecast to 2030},
  institution = {International Energy Agency},
  year        = {2024},
  note        = {\url{https://www.iea.org/reports/renewables-2024}}
}

@techreport{nrel2025ibr_cyberphysical,
  author      = {Reynolds, Tami},
  title       = {Cyber-Physical Challenges and Opportunities for Securing Inverter-Based Resources},
  institution = {National Renewable Energy Laboratory},
  year        = {2025},
  note        = {\url{https://docs.nlr.gov/docs/fy25osti/95985.pdf}}
}

@techreport{osti_1832209,
  author       = {Hupp, William and Saleem, Danish and Peterson, Jordan T. and Boyce, Kenneth},
  title        = {Cybersecurity Certification Recommendations for Interconnected Grid Edge Devices and Inverter Based Resources},
  institution  = {National Renewable Energy Laboratory},
  doi          = {10.2172/1832209},
  note         = {\url{https://www.osti.gov/biblio/1832209}},
  year         = {2021},
  }

@article{tuyen22,
  author={Tuyen, Nguyen Duc and Quan, Nguyen Sy and Linh, Vo Ba and Van Tuyen, Vu and Fujita, Goro},
  journal={IEEE Access}, 
  title={A Comprehensive Review of Cybersecurity in Inverter-Based Smart Power System Amid the Boom of Renewable Energy}, 
  year={2022},
  volume={10},
  number={},
  pages={35846-35875},
  doi={10.1109/ACCESS.2022.3163551}
  }

@ARTICLE{ahn24,
  author={Ahn, BoHyun and Kim, Taesic and Ahmad, Seerin and Mazumder, Sudip Kumar and Johnson, Jay and Mantooth, H. Alan and Farnell, Chris},
  journal={IEEE Trans. Power Electron.}, 
  title={An Overview of Cyber-Resilient Smart Inverters Based on Practical Attack Models}, 
  year={2024},
  volume={39},
  number={4},
  pages={4657-4673}
}

@techreport{nerc_ibr_report_2023,
  title       = {Inverter-Based Resource Performance Issues Report},
  institution = {North American Electric Reliability Corporation},
  year        = {2023},
  note        = {\url{https://www.nerc.com/globalassets/our-work/reports/white-papers/nerc_inverter-based_resource_performance_issues_public_report_2023.pdf}}
}

@INPROCEEDINGS{demarco98,
  author={DeMarco, C. L.},
  booktitle={31st Hawaii Int. Conf. Syst. Sci.}, 
  title={Design of predatory generation control in electric power systems}, 
  year={1998},
  volume={3},
  number={},
  pages={32-38},
}

@book{anderson03,
  title={Power System Control and Stability},
  author={Anderson, P. M. and Fouad, A. A.},
  year={2003},
  edition = {2nd},
  publisher={Wiley}
}

@book{saadat10,
  title={Power System Analysis},
  author={Saadat, H.},
  isbn={9780984543809},
  lccn={2010906755},
  year={2010},
  publisher={PSA Pub.}
}

@INPROCEEDINGS{demarco96,
  author={DeMarco, C. L. and Sariashkar, J. V. and Alvarado, F.},
  booktitle={IEEE Int. Conf. Control Appl.}, 
  title={The potential for malicious control in a competitive power systems environment}, 
  year={1996},
  volume={},
  number={},
  pages={462-467}
}

@ARTICLE{bahrani24,
  author={Bahrani, Behrooz and Ravanji, Mohammad Hasan and Kroposki, Benjamin and Ramasubramanian, Deepak and Guillaud, Xavier and Prevost, Thibault and Cutululis, Nicolaos-Antonio},
  journal={IEEE Power Energy Mag.}, 
  title={Grid-Forming Inverter-Based Resource Research Landscape: Understanding the Key Assets for Renewable-Rich Power Systems}, 
  year={2024},
  volume={22},
  number={2},
  pages={18-29}
}

@article{he21,
  title={Small-signal stability analysis for power system frequency regulation with renewable energy participation},
  author={He, Tingyi and Li, Shengnan and Wu, Shuijun and Li, Ke},
  journal={Math. Probl. Eng.},
  volume={2021},
  number={1},
  pages={5556062},
  year={2021},
  publisher={Wiley Online Library}
}

@ARTICLE{yang23,
  author={Yang, Ziqian and Zhan, Meng and Liu, Dan and Ye, Chang and Cao, Kan and Cheng, Shijie},
  journal={IEEE Trans. Power Syst.}, 
  title={Small-Signal Synchronous Stability of a New-Generation Power System With 100\% Renewable Energy}, 
  year={2023},
  volume={38},
  number={5},
  pages={4269-4280}
}

@INPROCEEDINGS{saber23,
  author={Saber, Ahmad Mohammad and Youssef, Amr and Svetinovic, Davor and Zeineldin, Hatem and El-Saadany, Ehab},
  booktitle={49th Annu. Conf. IEEE Ind. Electron. Soc.}, 
  title={Learning-Based Detection of Malicious {Volt-VAr} Control Parameters in Smart Inverters}, 
  year={2023},
  volume={},
  number={},
  pages={}
}

@ARTICLE{brown18,
  author={Brown, Hilary E. and Demarco, Christopher L.},
  journal={IEEE Trans. Smart Grid}, 
  title={Risk of Cyber-Physical Attack via Load With Emulated Inertia Control}, 
  year={2018},
  volume={9},
  number={6},
  pages={5854-5866}
}

@INPROCEEDINGS{roberts21,
  author={Roberts, Ciaran and Markovic, Uros and Arnold, Daniel and Callaway, Duncan S.},
  booktitle={IEEE Madrid PowerTech}, 
  title={Malicious Control of an Active Load in an Islanded Mixed-Source Microgrid}, 
  year={2021},
  pages={1-6}
}

@techreport{du23,
  title={Model specification of droop-controlled, grid-forming inverters {(REGFM\_A1)}},
  author={Du, Wei},
  year={2023},
  institution={Pacific Northwest National Laboratory},
  note = {\url{https://www.osti.gov/biblio/2229442}}
}

@manual{gurobi23,
  title  = {{Gurobi Optimizer Reference Manual}},
  author = {{Gurobi Optimization, LLC}},
  year   = {2023},
  note   = {\url{https://www.gurobi.com}}
}

@book{sauer98,
  title={Power System Dynamics and Stability},
  author={Sauer, P.W. and Pai, M.A.},
  isbn={9780136788300},
  lccn={97017360},
  url={https://books.google.ca/books?id=dO0eAQAAIAAJ},
  year={1998},
  publisher={Prentice Hall}
}

@inproceedings{subrahmanyan99,
  title={Eigenvector assignment},
  author={Subrahmanyan, Pradeep and Trumper, David},
  booktitle={Proc. {ACC}},
  volume={4},
  pages={2238--2243},
  year={1999},
  organization={IEEE}
}

@ARTICLE{klein77,
  author={Klein, G. and Moore, B.},
  journal={IEEE Trans. Autom. Control}, 
  title={Eigenvalue-generalized eigenvector assignment with state feedback}, 
  year={1977},
  volume={22},
  number={1},
  pages={140-141},
  doi={10.1109/TAC.1977.1101435}
}

@book{chen84,
  title={Linear System Theory and Design},
  author={Chen, Chi-Tsong},
  volume={301},
  year={1984},
  publisher={Holt, Rinehart and Winston New York}
}

@article{uijlings13,
  title={An independent analysis on the ability of Generators to ride through Rate of Change of Frequency values up to {2 Hz/s}},
  author={Uijlings, Willem and Street, DKLC and London, S},
  journal={EirGrid, London, UK, Rep},
  volume={16010927},
  year={2013}
}

@INPROCEEDINGS{amini15,
  author={Amini, Sajjad and Mohsenian-Rad, Hamed and Pasqualetti, Fabio},
  booktitle={IEEE PES Innov. Smart Grid Tech. Conf.}, 
  title={Dynamic load altering attacks in smart grid}, 
  year={2015},
  volume={},
  number={},
  pages={},
  doi={10.1109/ISGT.2015.7131791}
}

@article{vandewal01,
title = {A review of methods for input/output selection},
journal = {Automatica},
volume = {37},
number = {4},
pages = {487-510},
year = {2001},
issn = {0005-1098},
doi = {https://doi.org/10.1016/S0005-1098(00)00181-3},
url = {https://www.sciencedirect.com/science/article/pii/S0005109800001813},
author = {Marc {van de Wal} and Bram {de Jager}},
}

@INPROCEEDINGS{chanekar17,
  author={Chanekar, Prasad Vilas and Chopra, Nikhil and Azarm, Shapour},
  booktitle={Proc. {ACC}}, 
  title={Optimal actuator placement for linear systems with limited number of actuators}, 
  year={2017},
  volume={},
  number={},
  pages={334-339},
  doi={10.23919/ACC.2017.7962975},
  organization={IEEE}
}

@article{bamigbade2023cyberattack,
  title={Cyberattack on phase-locked loops in inverter-based energy resources},
  author={Bamigbade, Abdullahi and Dvorkin, Yury and Karri, Ramesh},
  journal={IEEE Trans. Smart Grid},
  volume={15},
  number={1},
  pages={821--833},
  year={2023}
}

@article{amini2016dynamic,
  title={Dynamic load altering attacks against power system stability: Attack models and protection schemes},
  author={Amini, Sajjad and Pasqualetti, Fabio and Mohsenian-Rad, Hamed},
  journal={IEEE Trans. Smart Grid},
  volume={9},
  number={4},
  pages={2862--2872},
  year={2016}
}

@article{olshevsky2014minimal,
  title={Minimal controllability problems},
  author={Olshevsky, Alex},
  journal={IEEE Control Netw. Syst.},
  volume={1},
  number={3},
  pages={249--258},
  year={2014},
  publisher={IEEE}
}

@article{summers2015submodularity,
  title={On submodularity and controllability in complex dynamical networks},
  author={Summers, Tyler H and Cortesi, Fabrizio L and Lygeros, John},
  journal={IEEE Control Netw. Syst.},
  volume={3},
  number={1},
  pages={91--101},
  year={2015},
  publisher={IEEE}
}

@inproceedings{katewa2021optimal,
  title={Optimal dynamic load-altering attacks against power systems},
  author={Katewa, Vaibhav and Pasqualetti, Fabio},
  booktitle={Proc. {ACC}},
  pages={4568--4573},
  year={2021},
  organization={IEEE}
}

@misc{zou2026,
  author = {Zou, Xiangyu},
  title = {Power System Dynamic Test Cases},
  year = {2026},
  publisher = {GitHub},
  howpublished = {\url{https://github.com/13zouhar/Power-System-Dynamic-Test-Cases}}
}
\end{document}